\documentclass[aps,prl,twocolumn,amsmath,10pt,superscriptaddress,floatfix,nofootinbib]{revtex4-1}

\usepackage{graphicx}
\usepackage{dcolumn}
\usepackage{bm}

\usepackage{multirow}

\usepackage[colorlinks=true,linktocpage=true,linkcolor=blue,citecolor=blue]{hyperref}

\graphicspath{{Fig/}}

\begin{document}

\title{The production of doubly charmed exotic hadrons in heavy ion collisions}

\author{Yuanyuan Hu}
\address{Guangdong Provincial Key Laboratory of Nuclear Science, Institute of Quantum Matter, South China Normal University, Guangzhou 510006, China}
\affiliation{Guangdong-Hong Kong Joint Laboratory of Quantum Matter, Southern Nuclear Science Computing Center, South China Normal University, Guangzhou 510006, China}

\author {Jinfeng Liao} \email{liaoji@indiana.edu}
\address{Physics Department and Center for Exploration of Energy and Matter, Indiana University, 2401 N Milo B. Sampson Lane, Bloomington, Indiana 47408, USA}

\author{Enke Wang}\email{wangek@scnu.edu.cn}
\address{Guangdong Provincial Key Laboratory of Nuclear Science, Institute of Quantum Matter, South China Normal University, Guangzhou 510006, China}
\affiliation{Guangdong-Hong Kong Joint Laboratory of Quantum Matter, Southern Nuclear Science Computing Center, South China Normal University, Guangzhou 510006, China}

\author{Qian Wang}\email{qianwang@m.scnu.edu.cn}
\address{Guangdong Provincial Key Laboratory of Nuclear Science, Institute of Quantum Matter, South China Normal University, Guangzhou 510006, China}
\affiliation{Guangdong-Hong Kong Joint Laboratory of Quantum Matter, Southern Nuclear Science Computing Center, South China Normal University, Guangzhou 510006, China}

\author{Hongxi Xing}\email{hxing@m.scnu.edu.cn}
\address{Guangdong Provincial Key Laboratory of Nuclear Science, Institute of Quantum Matter, South China Normal University, Guangzhou 510006, China}
\affiliation{Guangdong-Hong Kong Joint Laboratory of Quantum Matter, Southern Nuclear Science Computing Center, South China Normal University, Guangzhou 510006, China}

\author{Hui Zhang}\email{Mr.zhanghui@m.scnu.edu.cn}
\address{Guangdong Provincial Key Laboratory of Nuclear Science, Institute of Quantum Matter, South China Normal University, Guangzhou 510006, China}
\affiliation{Guangdong-Hong Kong Joint Laboratory of Quantum Matter, Southern Nuclear Science Computing Center, South China Normal University, Guangzhou 510006, China}

\date{\today}

\begin{abstract}
Hadron spectroscopy provides  direct physical measurements that shed light on the non-perturbative behavior of quantum chromodynamics (QCD). 
In particular, various exotic hadrons such as the newly  observed $T_{cc}^+$ by the LHCb collaboration,  
offer unique insights on the QCD dynamics in hadron structures. 
 In this letter, we demonstrate how heavy ion collisions can serve as a powerful venue
  for hadron spectroscopy study of  doubly charmed exotic hadrons
   by virtue of the extremely charm-rich environment created in such collisions.
    The yields  of $T_{cc}^+$ as well as its potential isospin partners 
 are computed within the molecular picture for Pb-Pb collisions
  at center-of-mass energy $2.76~\mathrm{TeV}$. 
  We find about three-order-of-magnitude enhancement
   in the production of $T_{cc}^+$ in Pb-Pb collisions
    as compared with the yield in proton-proton collisions,
     with a moderately smaller enhancement in the yields 
     of the isospin partners $T_{cc}^0$ and $T_{cc}^{++}$.
      The $T_{cc}^+$ yield is comparable to that of the $X(3872)$
       in the most central collisions while shows a considerably
        stronger decrease toward peripheral collisions, 
          due to a ``threshold'' effect of the required 
          double charm quarks for $T_{cc}^+$.  Final
           results for their rapidity and transverse momentum $p_T$ dependence
as well as the elliptic flow coefficient are reported and can be
 tested by future experimental measurements. 
\end{abstract}

\maketitle

{\it Introduction}~~
Because of the color confinement property of 
 Quantum chromodynamics (QCD), the theory of strong interaction,
experiments can only directly detect color singlet hadrons, instead of fundamental quarks and gluons.
As a result, the properties of hadrons, such as the origin of proton mass and spin, 
 structure functions, 
hadron spectroscopy,
 and hadron productions/decays in various processes,
 are very important for understanding the mystery of non-perturbative dynamics in QCD.   
 The study of hadron spectroscopy historically played crucial roles in the development of the conventional quark model, 
 with the classical example of the $\Omega$ baryon discovery that helped establish the model.  
 Today, extensive efforts on hadron spectroscopy have been actively carried out by
   experimental collaborations worldwide such as LHCb, 
   BESIII, BelleII, JLab, CMS, ATLAS, with particular interest
    in the search of possible exotic hadrons. 

Recently, the study of hadrons with two (or more) 
heavy quarks (or antiquarks) has attracted significant attention.
 Such states, while expected to exist in both conventional quark model and in the exotic sector, 
 are difficult to create and detect experimentally, 
 due to the apparent absence of any heavy-flavor valence quarks
  in the beam particles and thus highly suppressed production rate. 
  Nevertheless the available beam energy at the Large Hadron Collider (LHC) 
  and the highly capable detectors have started to offer such opportunity, 
  as shown in the observation of the $\Xi_{cc}^{++}$ by LHCb~\cite{LHCb:2017iph}. 
   Just earlier this month, the LHCb collaboration also reported a $J^P=1^+$ $T_{cc}^+$
 state with significance over $10~\sigma$ in the prompt $D^0D^0\pi^+$ invariant mass distribution 
 in the proton-proton ($pp$) collisions~\cite{LHCb:2021vvq, LHCb:2021auc},
  which is the first observation of a doubly charmed 
 tetraquark with quark content $cc\bar{u}\bar{d}$. 
 Its mass is very close to the $D^0D^{*+}$ and $D^+D^{*0}$
 thresholds with width about $410~\mathrm{keV}$. 
 There are many theoretical studies of 
 the open double heavy tetraquark system in the literature, 
 focusing on key issues such as  
 the formation mechanism (i.e. whether the double heavy tetraquark system is
 bound or not in either molecular picture or compact tetraquark picture)~\cite{Semay:1994ht,Pepin:1996id,Carlson:1987hh,Janc:2004qn,Vijande:2003ki,Lee:2009rt,Yang:2009zzp,Navarra:2007yw,Vijande:2007rf,Ebert:2007rn,Gelman:2002wf,Agaev:2021vur,Dong:2021bvy,Huang:2021urd,Li:2021zbw,Ren:2021dsi}, the double heavy exotic spectrum~\cite{Hudspith:2020tdf,Cheng:2020wxa,Qin:2020zlg,Drutskoy:2021euy,Xin:2021wcr,Chen:2021vhg,Weng:2021hje,Chen:2021tnn,Yang:2021zhe} and 
 the decay modes/production mechanism~\cite{Meng:2021jnw,Yan:2021wdl,Fleming:2021wmk,Jin:2021cxj}
  and their magnetic dipole moments~\cite{Azizi:2021aib}.  
 Detailed measurements on the $p_T$, rapidity, multiplicity and centrality dependence could
  help unravel the production mechanism and the internal structures of these  hadrons~\cite{Crkovska:2020tyr,LHCb:2019obz,Esposito:2020ywk,Braaten:2020iqw}. 
 The observation of the exotics with two or more heavy quarks 
 also provides a way to shed light on the potential symmetry
  (such as diquark-antiquark symmetry~\cite{Savage:1990di,Brambilla:2005yk,Fleming:2005pd}). 
  For reviews of the study of these exotic states, 
  we refer to Refs.~\cite{Chen:2016qju,Chen:2016spr,Dong:2017gaw,Lebed:2016hpi,Guo:2017jvc,Liu:2019zoy,Albuquerque:2018jkn,Yamaguchi:2019vea,Guo:2019twa,Brambilla:2019esw}. 

The crucial ``bottleneck'' for the creation of doubly charmed hadrons
 is the need of at least two charm quarks (which require production of
 two charm-anti-charm pairs in the initial hard scatterings). 
 In this regard, high energy heavy ion collisions  can serve
  as a powerful venue for the production of  doubly charmed exotic
   hadrons by virtue of the extremely charm-rich environment 
   in such collisions. Indeed, a central heavy ion collision at LHC
    energies could have many dozens of charm and anti-charm
     quarks available in a single event~\cite{Andronic:2003zv,Andronic:2017pug}.
      This unique advantage has been shown for the case of $X(3872)$ production
       in heavy ion collisions~\cite{Zhang:2020dwn,ExHIC:2010gcb,Chen:2021akx,CMS:2021znk,Wu:2020zbx,Fontoura:2019opw,Hong:2018mpk,Cho:2019syk,Esposito:2020ywk,Abreu:2020jsl,Albaladejo:2021cxj}. 
       It was proposed that the centrality dependence of $X(3872)$ yield could help distinguish a large size hadronic molecular scenario
 from a compact tetraquark scenario~\cite{Zhang:2020dwn}. 
 Given that the $T_{cc}^+$ production requires at least two 
 $c\bar c$ pairs while the  $X(3872)$ requires at least one pair,
  the heavy ion collision should be even more advantageous
   for producing the $T_{cc}^+$.  In this letter,
    we demonstrate this by computing the yields
     of $T_{cc}^+$ as well as its potential isospin partners $T_{cc}^{++}$ and $T_{cc}^0$
       in Pb-Pb collisions at center-of-mass energy $\sqrt{s}=2.76\,\rm{TeV}$. 
  The closeness of the $T_{cc}^+$ to the 
 $D^0D^{*+}$ and $D^+D^{*0}$ thresholds not only implies its potential molecular picture,
but also indicates large isospin breaking effects in its decay~\cite{Meng:2021jnw},
 which is similar to the case of the $X(3872)$~\cite{Gamermann:2009fv,Zhou:2017txt,Li:2012cs} 
  with the nearby $D^0\bar{D}^{*0}+c.c.$ and $D^+D^{*-}+c.c.$ thresholds.  
  Given the above, our calculation is performed within the molecular picture
   and we use the $X(3872)$ yield 
  to set a benchmark for the $T_{cc}^+$. 
  As we shall show below, 
  the yield of the $T_{cc}$~\footnote{When the charged property of $T_{cc}$ is not specified, it includes all the states, i.e. $T_{cc}^{++}$, $T_{cc}^0$ and $T_{cc}^+$.} is enhanced by roughly three-order-of-magnitude as compared
   with the yield in $pp$ collisions and is comparable to that 
   of the $X(3872)$ in the most central collisions while shows
    a considerably stronger decrease toward peripheral collisions. 
  Furthermore, we will also present results for the  rapidity
   and transverse momentum $p_T$ dependence
as well as the elliptic flow coefficient that can be tested by future measurements. 

{\it Framework}~~
For this study, we adopt the framework developed in Ref.~\cite{Zhang:2020dwn} 
to  generate a total of one million minimum bias events from the default version
 of AMPT transport model for Pb-Pb collisions at $\sqrt{s}=2.76\,\rm{TeV}$ 
 for simulating the production of both $X(3872)$ and the $T_{cc}$ in these collisions. 
The charmed mesons $D$ and $D^*$ ($\bar{D}^*$) are collected after the hadronization process
and coalesced to the $T_{cc}$ states  with the following 
 conditions based on the molecular picture: the relative distance within the region $[5 \rm{fm},7 \rm{fm}]$ and
   the invariant mass within the region $[2M_D,2M_{D^*}]$.
   In the $DD^*$ molecular picture, there are  four possible states:
   \begin{eqnarray}
  T_{cc}^0:&\quad& D^0D^{*0}\quad I=1,\quad I_3=-1,\\
  T_{cc}^{++}: &\quad& D^+D^{*+} \quad I=1,\quad I_3=1,\\
   T_{cc}^{(\prime)+} :&\quad& D^{0/+}D^{*+/0} \quad I=1,0\quad I_3=0.
   \end{eqnarray}
   The first two correspond to the iso-triplet states $T_{cc}^0$ and $T_{cc}^{++}$ 
   ~\cite{Junnarkar:2018twb,Faustov:2021hjs,Yang:2009zzp} that may potentially be produced.  
   The last two  $T_{cc}^+$ and $T_{cc}^{\prime +}$ could be mixtures of isospin triplet and singlet. 
As the possible interference between the $D^0D^{*+}$
   and $D^+D^{*0}$ components is not implemented in the simulation framework,
    we do not distinguish these $T_{cc}^{(\prime)+}$ components here.  
    In what follows, we use $T_{cc}^+$ to denote these two states.
   As the charmed mesons are formed in the AMPT model based on quark flavor content while lacking spin information, 
   the relative yield ratios between e.g. $D^{*+}$ versus $D^+$ or that between $D^{*0}$ and $D^0$ 
    need to be estimated from  the thermal model relation
    \begin{equation}
R(\frac{A}{B})\equiv \frac{\text{Yield}(A)}{\text{Yield}(B)}= e^{-(m_A-m_B)/T_{\text{freezeout}}},
\label{eq:thermal}
\end{equation}
with $m_A$ and $m_B$ the masses of hadrons A and B, respectively. 
Here $T_{\text{freezeout}}\simeq160~\mathrm{MeV}$ is the freeze-out temperature. 
With the physical masses of $D^{(*)+}$ and $D^{(*)0}$, 
we find that the relevant fractions to be  $29.3\%$ versus $70.7\%$ for $D^{*+}$ 
versus $D^+$ and  $29.2\%$ versus $70.8\%$ for $D^{*0}$ versus $D^0$, respectively. 
 To calibrate potential influence associated with this procedure, 
 we estimate the uncertainty of our results by varying the fractions in the regions $[20\%,40\%]$
and $[80\%,60\%]$ for $D^+$ and $D^{*+}$, respectively. 
As a sanity check, we also verified that our model simulation results for the total $D+D^*$ yields 
agree with experimental measurements~\cite{ALICE:2015vxz}. 
   
{\it Results and Discussions} 
In this work we focus on estimating the $T_{cc}^+$ yield  from the coalescence of  $D^0D^{*+}$
and $D^+D^{*0}$ pairs within the aforementioned framework.
 As a benchmark for comparison, we also estimate the $X(3872)$ production within the same framework 
 as the average yield from coalescence of the 
$D^0\bar{D}^{0*}$, $D^{0*}\bar{D}^0$, $D^+D^{-*}$, $D^{+*}D^-$
pairs~\cite{Zhang:2020dwn}. Additionally   
the yields of $T_{cc}^0$ and $T_{cc}^{++}$ states are computed from coalescence of the $D^0D^{*0}$
and $D^+D^{*+}$ pairs, respectively. 
With a total of one million minimum bias events from our simulations for Pb-Pb collisions at $
\sqrt{s}=2.76~\mathrm{TeV}$, the inclusive yields of the $X(3872)$, $T_{cc}^0$, $T_{cc}^{++}$
and $T_{cc}^+$ are found to be around $49000$, $44000$, $44000$ and $50000$, respectively, for
$\mathrm{R}(\frac{D}{D+D^*})=70\%$.
 The fact that these four are almost of the same order may appear counter-intuitive at first sight.  
 Given that the $c$ and $\bar{c}$ quarks must be pair produced
  and thus have the same abundance in each event,
   a naive counting may suggest that there would be more 
 $c\bar{c}$ pairs than $cc$ pairs and thus more likelihood to
  form $X(3872)$ than $T_{cc}$. Indeed, assuming on average
   there are $N$ charm and $N$ anti-charm quarks generated in a given event,
    there would be a total of $N^2$  $c\bar{c}$ pairs and
     $N*(N-1)/2$ $cc$ pairs.   For $N>>1$, roughly it is a factor of 2 difference 
     which is also confirmed from our simulations. However, the formation of
      either $X(3872)$ or $T_{cc}^+$ in the molecular picture requires a
       $D$ with a ${D}^*$ instead of the charm quarks/anti-quarks. 
       This changes the counting: assuming a ratio $\mathrm{R}$ for $\frac{D}{D+D^*}$ 
       (and similarly for $\frac{\bar{D}}{\bar{D}+\bar{D}^*}$), there would be roughly $N R$ of $D$ and $N(1-R)$ of $D^*$
        as well as $NR$ of $\bar{D}$ and $N(1-R)$ of $\bar{D}^*$. 
        So in the end one gets a similar count of $N^2 R(1-R)$ for 
        both $DD^*$ pairs and $D\bar{D}^*$ pairs~\footnote{Notice that the yields of the $X(3872)$ ($49000$ discussed above) is equal to that of the $T_{cc}^+$ ($50000$ discussed above) within their statistic uncertainties.}. This helps explain why the inclusive yields of them are fairly close
        at large $N$, i.e. the central centrality region discussed below. 

\begin{figure}[hbt!] 
\includegraphics[height=15cm]{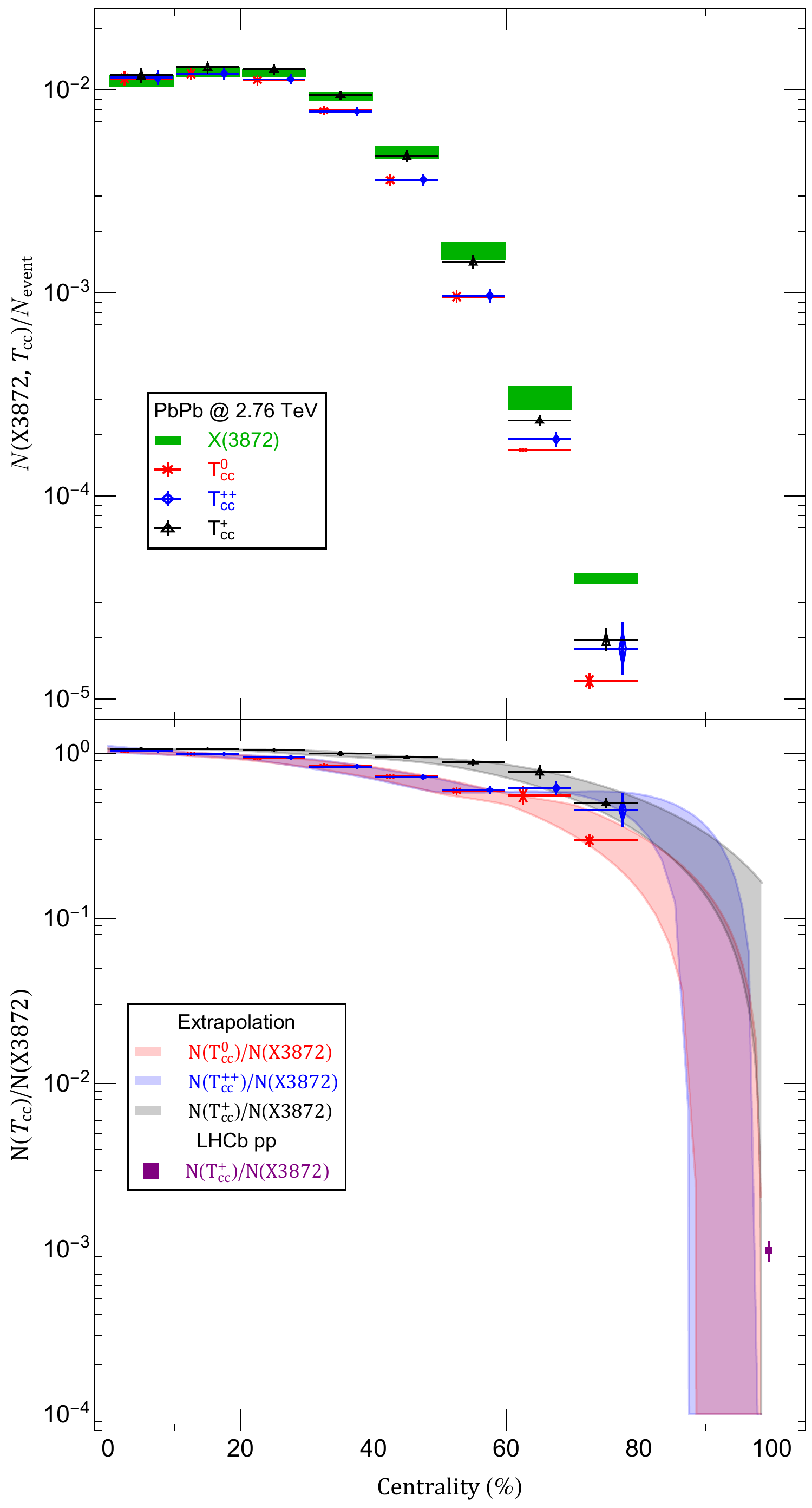}
\caption
{ The centrality dependence of the $X(3872)$ (green solid boxes), 
$T_{cc}^0$ (red stars), $T_{cc}^{++}$ (blue diamonds) and
$T_{cc}^+$ (black triangles) in the $D\bar{D}^*+c.c.$, $D^0D^{*0}$, $D^+D^{*+}$ and $D^0D^{*+}/D^+D^{*0}$ 
hadronic molecular picture, respectively. The bands reflect the uncertainty due to
constituent composition as discussed in Eq.~\eqref{eq:thermal} that are obtained from varying the
composition fraction by $\pm10\%$. The ratios of the yields for the three
 $T_{cc}$s relative to that for the $X(3872)$ are also presented in the lower panel,
  where an extrapolation with  a third-order polynomial function
   of the $T_{cc}^0$ (gray shaded band) and $T_{cc}^{++}$ (pink shaded band) yield
    ratios toward ultra-peripheral region are also presented. 
    The purple square is the ratio extracted from the experimental data~\cite{LHCb:2021auc,LHCb:2021vvq, LHCb:2011zzp,CMS:2013fpt,LHCb:2019obz,LHCb:2020fvo,LHCb:2020sey,Heijn:2728971}.} 
 \label{fig:centrality}
\end{figure}

To see the fireball volume effect on the $T_{cc}$ production, 
we plot the centrality dependence of their yields in Fig.~\ref{fig:centrality}, 
where a  significant  decrease from central to peripheral collisions is found. 
This trend may be expected for 
a  hadron molecule with relatively large size.
 In heavy ion collisions the  charm quarks and  anti-quarks
are carried by bulk flow and the produced charm mesons 
spread out over the whole fireball~\cite{Zhang:2020dwn}.
 In peripheral collisions the fireball volume 
becomes small and results in a relatively 
small spatial separation between the relevant charm mesons, 
which disfavors the formation of molecular states.  
Our results for centrality dependence clearly demonstrate
 the unparalleled advantage of heavy ion collisions
  for producing the doubly charmed exotic hadrons,
   especially in the central and semi-central collisions. 

Furthermore, a comparison between the  $T_{cc}$  yields
 and $X(3872)$ yield as shown by the ratio of the two
  in Fig.~\ref{fig:centrality} (lower panel) reveals
   an even stronger suppression of the former in the peripheral collisions.  
This behavior points to an interesting   ``threshold'' effect of the required
 double charm quarks for $T_{cc}$ formations. Again, let us assume
  an average of  $N$ charm and $N$ anti-charm quarks in a given event with  
  a ratio $\mathrm{R}$ for $\frac{D}{D+D^*}$ (and similarly for $\frac{\bar{D}}{\bar{D}+\bar{D}^*}$). 
  The production of $X(3872)$ requires at least one pair of $D$+$\bar{D}^*$ or $\bar{D}$+$D^*$, 
  for which the probability is $P_X=1-R^{2N}-(1-R)^{2N}$. 
  The production of $T_{cc}$, on the other hand, requires at least one pair of $D$+${D}^*$, 
  for which the probability is $P_T=1-R^{N}-(1-R)^{N}$. 
  Note that $R<1$ and $(1-R)<1$, so the chance becomes
   considerably smaller for $T_{cc}$ production than $X(3872)$ 
   especially when the number $N$ becomes smaller. 
   To given an extreme example: when $N\to 1$ (i.e. in the limit of only one $c\bar{c}$ pair per event), 
   $P_T =0$ while $P_X>0$. The essence of such a suppression on the $T_{cc}$ production
    is essentially a ``threshold" effect occurring in the limit of ultra-low charm abundance.
     In heavy ion collisions, the number $N$ of $c\bar{c}$ pairs per event scales
      with the number of initial hard scatterings which in turn scales
       with the so-called binary collision number $N_{coll}$~\cite{Loizides:2017ack}. 
       The $N_{coll}$ drops very rapidly from central toward peripheral region, 
       thus providing an explanation of the observed pattern for $T_{cc}$ centrality dependence. 

To gain further insight,  we perform an extrapolation of the centrality dependence 
with a third-order polynomial function for the relative yield ratio between $T_{cc}$ and $X(3872)$
 toward the ultra-peripheral regime, as indicated by the color bands in Fig.~\ref{fig:centrality} (lower panel).
One can see that the yield of the $T_{cc}^+$ 
is at least three orders smaller than that of the $X(3872)$ in
 the ultra-peripheral collisions, which are expected to approach the limit of elementary $pp$ collisions.
   We note that the extrapolated result in that limit shows consistency with the corresponding
    ratio between  $T_{cc}^+$ and $X(3872)$ in $pp$ collisions from LHCb measurements
    ~\cite{LHCb:2021auc,LHCb:2021vvq, LHCb:2011zzp,CMS:2013fpt,LHCb:2019obz,LHCb:2020fvo,LHCb:2020sey,Heijn:2728971}.  
Finally, the extrapolation suggests  the yield of the iso-triplet states is at least
 two orders of magnitude smaller than that of the $T_{cc}^+$,
which may provide a plausible reason for the absence of
 the iso-triplet $T_{cc}^{++}$ so far in LHCb data~\cite{LHCb:2021auc}. 

\begin{figure}[hbt!] 
\includegraphics[height=8cm]{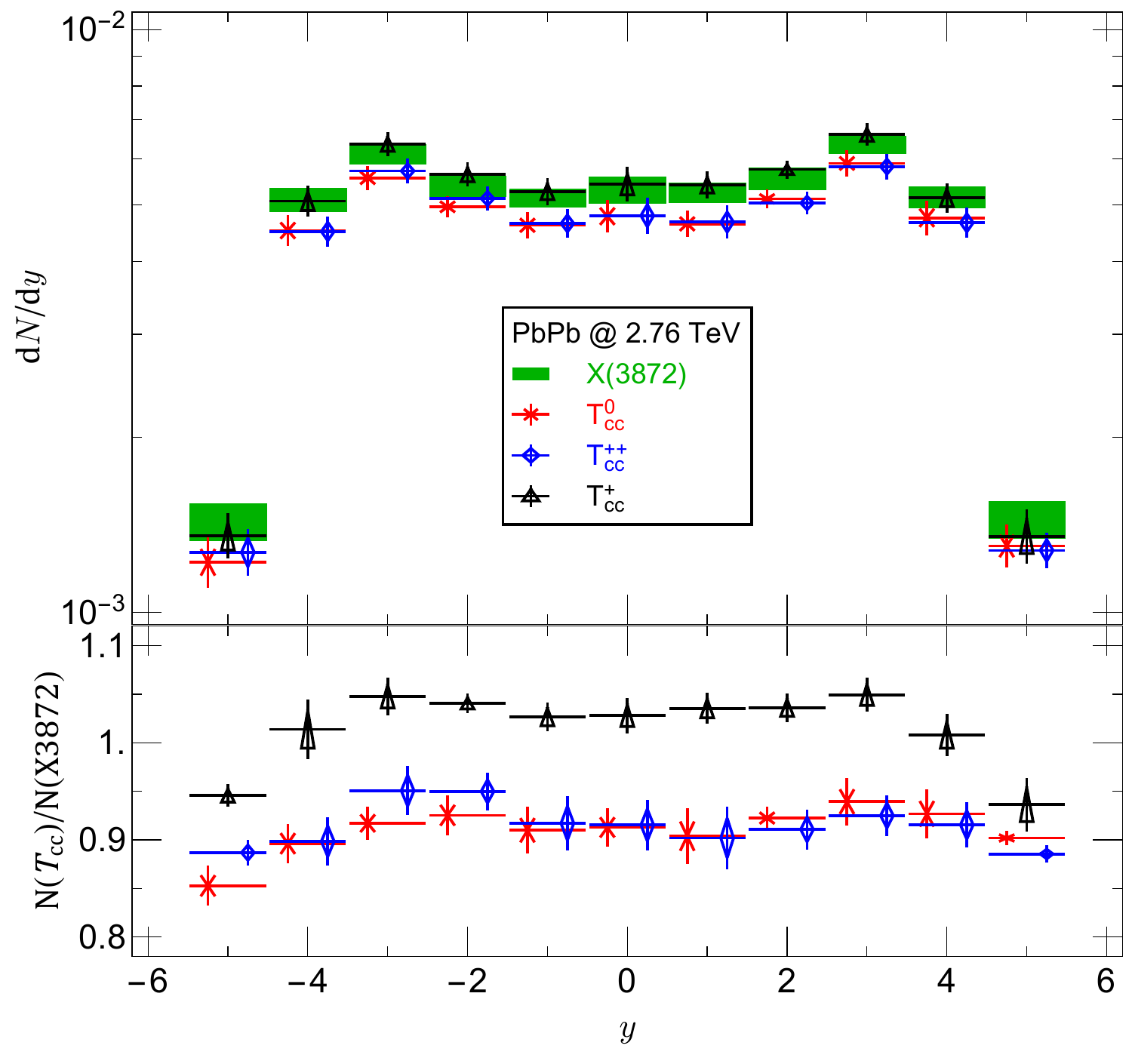}
\includegraphics[height=8cm]{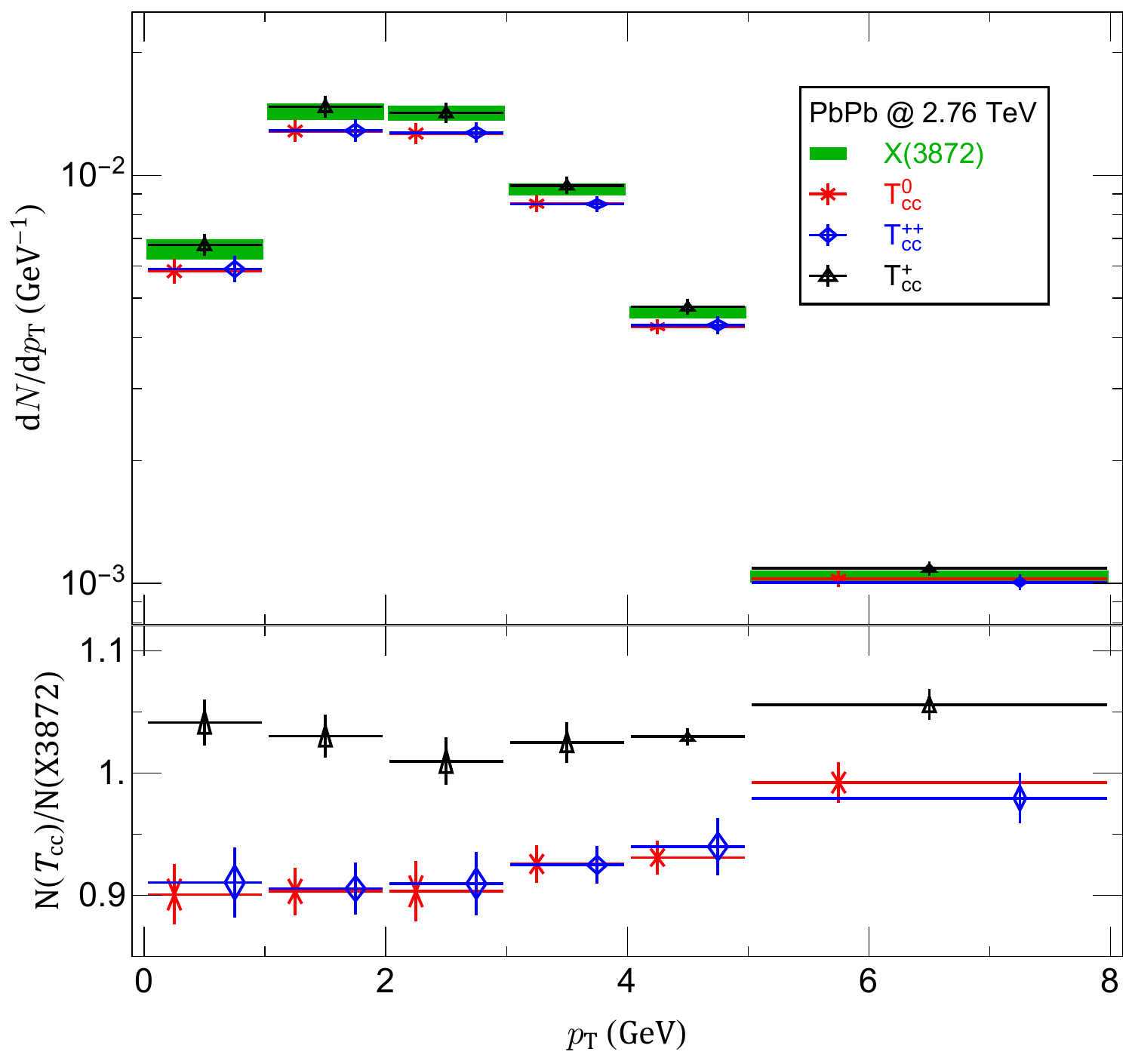}
\caption
{ The rapidity and transverse momentum distributions for the $X(3872)$ (green solid boxes),
 $T_{cc}^0$ (red stars), $T_{cc}^{++}$ (blue diamonds) and
$T_{cc}^+$ (black triangles) in the $D\bar{D}^*+c.c.$, $D^0D^{*0}$, $D^+D^{*+}$ and $D^0D^{*+}/D^+D^{*0}$ 
hadronic molecular picture, respectively. 
The uncertainties are obtained in the same way as that in Fig.~\ref{fig:centrality}.
 The ratios of the yields for the three
 $T_{cc}$s relative to that for the $X(3872)$ are also presented in the lower panels.  } 
 \label{fig:rapidity}
\end{figure}

\begin{figure}[hbt!] 
    \includegraphics[height=5cm]{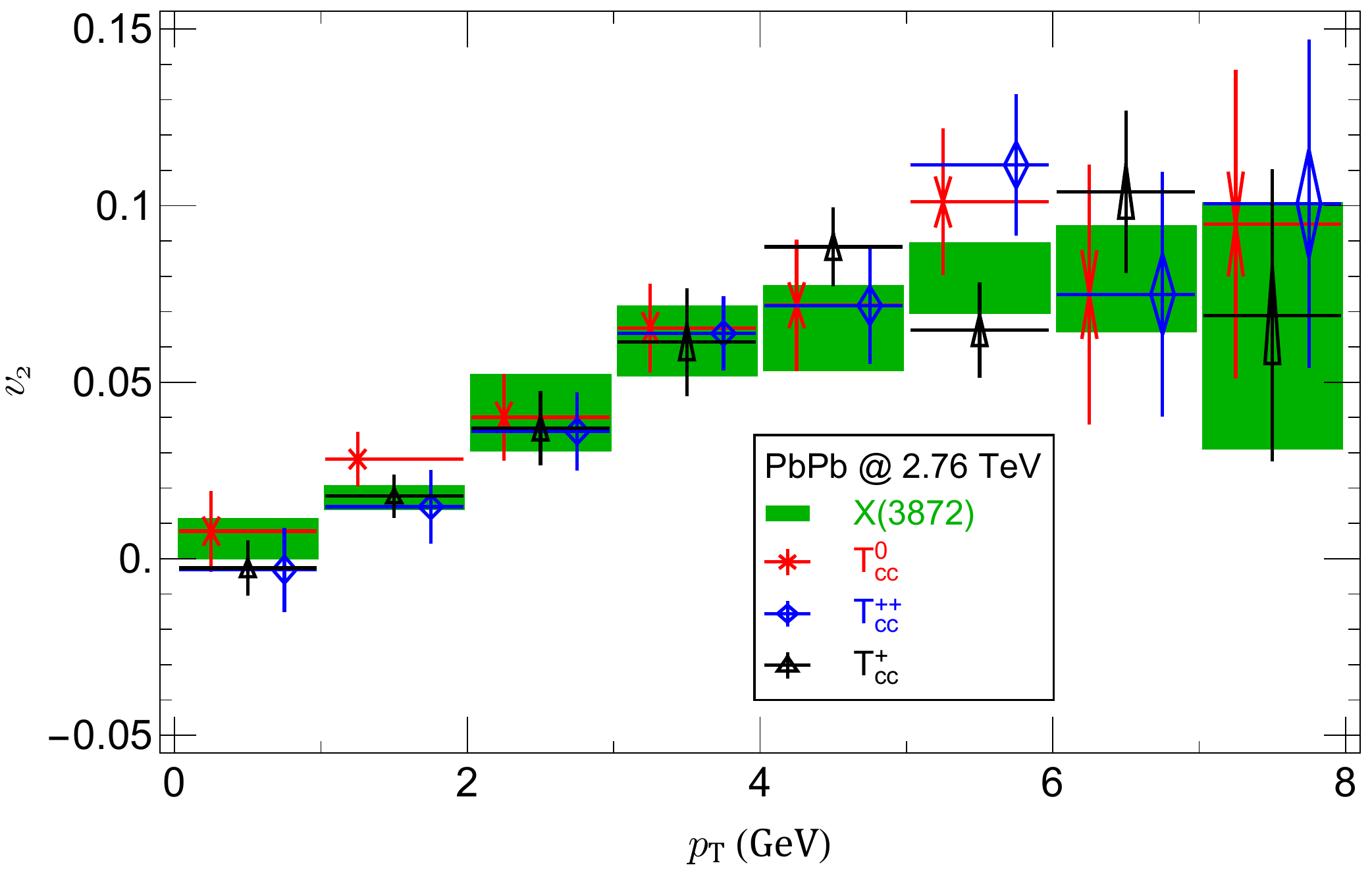}
    \caption
    { The elliptic flow coefficient $v_2$ versus transverse momentum $p_T$ for
     the $X(3872)$ (green solid boxes), $T_{cc}^0$ (red stars), $T_{cc}^{++}$ (blue diamonds) and
$T_{cc}^+$ (black triangles) in the $D\bar{D}^*+c.c.$, $D^0D^{*0}$, $D^+D^{*+}$ and $D^0D^{*+}/D^+D^{*0}$ 
hadronic molecular picture, respectively. The uncertainties are obtained in the same way as that in Fig.~\ref{fig:centrality}.
 } 
 \label{fig:elliptic}
\end{figure}

In Fig.~\ref{fig:rapidity} we present the rapidity and the transverse momentum 
distributions of these states, which are found to be similar to those 
of the usual hadrons~\cite{CMS:2011aqh,ALICE:2013jfw}. 
The rapidity dependence is flat in the middle and decreasing at the  
forward/backward region. The $p_T$ spectra decreases very strongly with increasing 
$p_T$, which may be expected from  production from the thermal source with radial flow~\cite{Zhang:2020dwn}.  
Finally we also show the results for the elliptic flow coefficient $v_2$ of these states
 in Fig.~\ref{fig:elliptic}, which suggest a very similar elliptic flow pattern among these states.
  The elliptic flow of a particle like $X(3872)$ or $T_{cc}$ would be sensitive
   to the charm mesons that coalesce into them, especially the spatial distributions
    of these mesons in the fireball. The similarity in $v_2$ among them is
    due to a similar spatial distributions of various  $D$, $D^*$, $\bar{D}$
     and $\bar{D}^*$ mesons, as we verified from our simulations.

{\it Summary and Outlook}~~
In this work, we estimate the yields of the recently observed $T_{cc}^+$
as well as its potential isospin partners $T_{cc}^0$ and $T_{cc}^{++}$
within the $DD^*$ hadronic molecular picture in Pb-Pb collisions
 at center-of-mass energy $2.76~\mathrm{TeV}$. 
 Our main finding is a strong enhancement, about three orders of magnitude, 
 for the $T_{cc}$ yield in the central collisions as compared with the
  very peripheral collisions which would approach the $pp$ baseline.
   In comparison with the $X(3872)$ yield computed in the same framework,
    we find their inclusive yields are close to each other in the central region
    but the $T_{cc}$ production shows a much stronger suppression into the peripheral region, 
    which can be understood from an interesting ``threshold" effect of 
    the required double charm quarks for $T_{cc}$ formation. 
    Final results are obtained for the rapidity and transverse momentum $p_T$
     dependence of $T_{cc}$ production 
as well as for the elliptic flow coefficient. 
Overall, we have demonstrated how heavy ion collisions
 can serve as a powerful venue for hadron spectroscopy study of  
 doubly charmed exotic hadrons by virtue of the extremely 
 charm-rich environment created in such collisions.   
 It would be exciting to anticipate future experimental efforts
  that will look for $T_{cc}$ states in heavy ion collisions
   and test the findings from the present work. 
   Given the advantage of heavy ion collisions
    in producing an abundance of these doubly charmed exotics, 
    it is conceivable that measurements from heavy ion experiments
     would offer great opportunities to nail down their structures and properties.

{\it Acknowledgements}~~
Discussions with Qingnian Xu are acknowledged. 
This work is partly supported by Guangdong Major Project of Basic and Applied Basic Research No.~2020B0301030008,
the National Natural Science Foundation of China with Grant No.~12035007, No.~12047523, NO.~12022512, 
Science and Technology Program of Guangzhou No.~2019050001,
Guangdong Provincial funding with Grant No.~2019QN01X172.
Q.W. is also supported by the NSFC and the Deutsche Forschungsgemeinschaft (DFG, German
Research Foundation) through the funds provided to the Sino-German Collaborative
Research Center TRR110 ``Symmetries and the Emergence of Structure in QCD"
(NSFC Grant No. 12070131001, DFG Project-ID 196253076-TRR 110). 
J.L. is supported by the U.S. NSF Grant No. PHY-1913729.



\begin{thebibliography}{}
\bibitem{LHCb:2017iph}
R.~Aaij \textit{et al.} [LHCb],
Phys. Rev. Lett. \textbf{119}, no.11, 112001 (2017)
[arXiv:1707.01621 [hep-ex]].

\bibitem{LHCb:2021vvq}
R.~Aaij \textit{et al.} [LHCb],
[arXiv:2109.01038 [hep-ex]].

\bibitem{LHCb:2021auc}
R.~Aaij \textit{et al.} [LHCb],
[arXiv:2109.01056 [hep-ex]].

\bibitem{Semay:1994ht}
C.~Semay and B.~Silvestre-Brac,
Z. Phys. C \textbf{61}, 271-275 (1994)

\bibitem{Pepin:1996id}
S.~Pepin, F.~Stancu, M.~Genovese and J.~M.~Richard,
Phys. Lett. B \textbf{393}, 119-123 (1997)
[arXiv:hep-ph/9609348 [hep-ph]].

\bibitem{Carlson:1987hh}
J.~Carlson, L.~Heller and J.~A.~Tjon,
Phys. Rev. D \textbf{37}, 744 (1988)

\bibitem{Janc:2004qn}
D.~Janc and M.~Rosina,
Few Body Syst. \textbf{35}, 175-196 (2004)
[arXiv:hep-ph/0405208 [hep-ph]].

\bibitem{Vijande:2003ki}
J.~Vijande, F.~Fernandez, A.~Valcarce and B.~Silvestre-Brac,
Eur. Phys. J. A \textbf{19}, 383 (2004)
[arXiv:hep-ph/0310007 [hep-ph]].

\bibitem{Lee:2009rt}
S.~H.~Lee and S.~Yasui,
Eur. Phys. J. C \textbf{64}, 283-295 (2009)
[arXiv:0901.2977 [hep-ph]].

\bibitem{Yang:2009zzp}
Y.~Yang, C.~Deng, J.~Ping and T.~Goldman,
Phys. Rev. D \textbf{80}, 114023 (2009)

\bibitem{Navarra:2007yw}
F.~S.~Navarra, M.~Nielsen and S.~H.~Lee,
Phys. Lett. B \textbf{649}, 166-172 (2007)
[arXiv:hep-ph/0703071 [hep-ph]].

\bibitem{Vijande:2007rf}
J.~Vijande, E.~Weissman, A.~Valcarce and N.~Barnea,
Phys. Rev. D \textbf{76}, 094027 (2007)
[arXiv:0710.2516 [hep-ph]].

\bibitem{Ebert:2007rn}
D.~Ebert, R.~N.~Faustov, V.~O.~Galkin and W.~Lucha,
Phys. Rev. D \textbf{76}, 114015 (2007)
[arXiv:0706.3853 [hep-ph]].

\bibitem{Gelman:2002wf}
B.~A.~Gelman and S.~Nussinov,
Phys. Lett. B \textbf{551}, 296-304 (2003)
[arXiv:hep-ph/0209095 [hep-ph]].

\bibitem{Agaev:2021vur}
S.~S.~Agaev, K.~Azizi and H.~Sundu,
[arXiv:2108.00188 [hep-ph]].

\bibitem{Dong:2021bvy}
X.~K.~Dong, F.~K.~Guo and B.~S.~Zou,
[arXiv:2108.02673 [hep-ph]].

\bibitem{Huang:2021urd}
Y.~Huang, H.~Q.~Zhu, L.~S.~Geng and R.~Wang,
[arXiv:2108.13028 [hep-ph]].

\bibitem{Li:2021zbw}
N.~Li, Z.~F.~Sun, X.~Liu and S.~L.~Zhu,
Chin. Phys. Lett. \textbf{38}, 092001 (2021)
[arXiv:2107.13748 [hep-ph]].

\bibitem{Ren:2021dsi}
H.~Ren, F.~Wu and R.~Zhu,
[arXiv:2109.02531 [hep-ph]].

\bibitem{Hudspith:2020tdf}
R.~J.~Hudspith, B.~Colquhoun, A.~Francis, R.~Lewis and K.~Maltman,
Phys. Rev. D \textbf{102}, 114506 (2020)
[arXiv:2006.14294 [hep-lat]].

\bibitem{Cheng:2020wxa}
J.~B.~Cheng, S.~Y.~Li, Y.~R.~Liu, Z.~G.~Si and T.~Yao,
Chin. Phys. C \textbf{45}, no.4, 043102 (2021)
[arXiv:2008.00737 [hep-ph]].

\bibitem{Qin:2020zlg}
Q.~Qin, Y.~F.~Shen and F.~S.~Yu,
Chin. Phys. C \textbf{45}, 103106 (2021)
[arXiv:2008.08026 [hep-ph]].

\bibitem{Drutskoy:2021euy}
A.~Drutskoy,
[arXiv:2101.09891 [hep-ph]].

\bibitem{Xin:2021wcr}
Q.~Xin and Z.~G.~Wang,
[arXiv:2108.12597 [hep-ph]].

\bibitem{Chen:2021vhg}
R.~Chen, Q.~Huang, X.~Liu and S.~L.~Zhu,
[arXiv:2108.01911 [hep-ph]].

\bibitem{Weng:2021hje}
X.~Z.~Weng, W.~Z.~Deng and S.~L.~Zhu,
[arXiv:2108.07242 [hep-ph]].

\bibitem{Chen:2021tnn}
X.~Chen,
[arXiv:2109.02828 [hep-ph]].

\bibitem{Yang:2021zhe}
G.~Yang, J.~Ping and J.~Segovia,
[arXiv:2109.04311 [hep-ph]].

\bibitem{Meng:2021jnw}
L.~Meng, G.~J.~Wang, B.~Wang and S.~L.~Zhu,
[arXiv:2107.14784 [hep-ph]].

\bibitem{Yan:2021wdl}
M.~J.~Yan and M.~P.~Valderrama,
[arXiv:2108.04785 [hep-ph]].

\bibitem{Fleming:2021wmk}
S.~Fleming, R.~Hodges and T.~Mehen,
[arXiv:2109.02188 [hep-ph]].

\bibitem{Jin:2021cxj}
Y.~Jin, S.~Y.~Li, Y.~R.~Liu, Q.~Qin, Z.~G.~Si and F.~S.~Yu,
[arXiv:2109.05678 [hep-ph]].

\bibitem{Azizi:2021aib}
K.~Azizi and U.~\"Ozdem,
[arXiv:2109.02390 [hep-ph]].

\bibitem{Crkovska:2020tyr}
J.~Crkovska [LHCb],
PoS \textbf{LHCP2020}, 173 (2021)

\bibitem{LHCb:2019obz}
 [LHCb],
LHCb-CONF-2019-005.

\bibitem{Esposito:2020ywk}
A.~Esposito, E.~G.~Ferreiro, A.~Pilloni, A.~D.~Polosa and C.~A.~Salgado,
Eur. Phys. J. C \textbf{81}, 669 (2021)
[arXiv:2006.15044 [hep-ph]].

\bibitem{Braaten:2020iqw}
E.~Braaten, L.~P.~He, K.~Ingles and J.~Jiang,
Phys. Rev. D \textbf{103}, no.7, L071901 (2021)
[arXiv:2012.13499 [hep-ph]].

\bibitem{Savage:1990di}
M.~J.~Savage and M.~B.~Wise,
Phys. Lett. B \textbf{248}, 177-180 (1990)

\bibitem{Brambilla:2005yk}
N.~Brambilla, A.~Vairo and T.~Rosch,
Phys. Rev. D \textbf{72}, 034021 (2005)
[arXiv:hep-ph/0506065 [hep-ph]].

\bibitem{Fleming:2005pd}
S.~Fleming and T.~Mehen,
Phys. Rev. D \textbf{73}, 034502 (2006)
[arXiv:hep-ph/0509313 [hep-ph]].

\bibitem{Chen:2016qju}
H.~X.~Chen, W.~Chen, X.~Liu and S.~L.~Zhu,
Phys. Rept. \textbf{639}, 1-121 (2016)
[arXiv:1601.02092 [hep-ph]].

\bibitem{Chen:2016spr}
H.~X.~Chen, W.~Chen, X.~Liu, Y.~R.~Liu and S.~L.~Zhu,
Rept. Prog. Phys. \textbf{80}, no.7, 076201 (2017)
[arXiv:1609.08928 [hep-ph]].

\bibitem{Dong:2017gaw}
Y.~Dong, A.~Faessler and V.~E.~Lyubovitskij,
Prog. Part. Nucl. Phys. \textbf{94}, 282-310 (2017)

\bibitem{Lebed:2016hpi}
R.~F.~Lebed, R.~E.~Mitchell and E.~S.~Swanson,
Prog. Part. Nucl. Phys. \textbf{93}, 143-194 (2017)
[arXiv:1610.04528 [hep-ph]].

\bibitem{Guo:2017jvc}
F.~K.~Guo, C.~Hanhart, U.-G.~Mei\ss{}ner, Q.~Wang, Q.~Zhao and B.~S.~Zou,
Rev. Mod. Phys. \textbf{90}, no.1, 015004 (2018)
[arXiv:1705.00141 [hep-ph]].

\bibitem{Liu:2019zoy}
Y.~R.~Liu, H.~X.~Chen, W.~Chen, X.~Liu and S.~L.~Zhu,
Prog. Part. Nucl. Phys. \textbf{107}, 237-320 (2019)
[arXiv:1903.11976 [hep-ph]].

\bibitem{Albuquerque:2018jkn}
R.~M.~Albuquerque, J.~M.~Dias, K.~P.~Khemchandani, A.~Mart\'\i{}nez Torres, F.~S.~Navarra, M.~Nielsen and C.~M.~Zanetti,
J. Phys. G \textbf{46}, no.9, 093002 (2019)
[arXiv:1812.08207 [hep-ph]].

\bibitem{Yamaguchi:2019vea}
Y.~Yamaguchi, A.~Hosaka, S.~Takeuchi and M.~Takizawa,
J. Phys. G \textbf{47}, no.5, 053001 (2020)
[arXiv:1908.08790 [hep-ph]].

\bibitem{Guo:2019twa}
F.~K.~Guo, X.~H.~Liu and S.~Sakai,
Prog. Part. Nucl. Phys. \textbf{112}, 103757 (2020)
[arXiv:1912.07030 [hep-ph]].

\bibitem{Brambilla:2019esw}
N.~Brambilla, S.~Eidelman, C.~Hanhart, A.~Nefediev, C.~P.~Shen, C.~E.~Thomas, A.~Vairo and C.~Z.~Yuan,
Phys. Rept. \textbf{873}, 1-154 (2020)
[arXiv:1907.07583 [hep-ex]].

\bibitem{Andronic:2003zv}
A.~Andronic, P.~Braun-Munzinger, K.~Redlich and J.~Stachel,
Phys. Lett. B \textbf{571}, 36-44 (2003)
[arXiv:nucl-th/0303036 [nucl-th]].

\bibitem{Andronic:2017pug}
A.~Andronic, P.~Braun-Munzinger, K.~Redlich and J.~Stachel,
Nature \textbf{561}, no.7723, 321-330 (2018)
[arXiv:1710.09425 [nucl-th]].

\bibitem{Zhang:2020dwn}
H.~Zhang, J.~Liao, E.~Wang, Q.~Wang and H.~Xing,
Phys. Rev. Lett. \textbf{126}, no.1, 012301 (2021)
[arXiv:2004.00024 [hep-ph]].

\bibitem{ExHIC:2010gcb}
S.~Cho \textit{et al.} [ExHIC],
Phys. Rev. Lett. \textbf{106}, 212001 (2011)
[arXiv:1011.0852 [nucl-th]].

\bibitem{Chen:2021akx}
B.~Chen, L.~Jiang, X.~H.~Liu, Y.~Liu and J.~Zhao,
[arXiv:2107.00969 [hep-ph]].

\bibitem{CMS:2021znk}
A.~M.~Sirunyan \textit{et al.} [CMS],
[arXiv:2102.13048 [hep-ex]].

\bibitem{Wu:2020zbx}
B.~Wu, X.~Du, M.~Sibila and R.~Rapp,
Eur. Phys. J. A \textbf{57}, no.4, 122 (2021)
[arXiv:2006.09945 [nucl-th]].

\bibitem{Fontoura:2019opw}
C.~E.~Fontoura, G.~Krein, A.~Valcarce and J.~Vijande,
Phys. Rev. D \textbf{99}, no.9, 094037 (2019)
[arXiv:1905.03877 [hep-ph]].

\bibitem{Hong:2018mpk}
J.~Hong, S.~Cho, T.~Song and S.~H.~Lee,
Phys. Rev. C \textbf{98}, no.1, 014913 (2018)
[arXiv:1804.05336 [nucl-th]].

\bibitem{Cho:2019syk}
S.~Cho and S.~H.~Lee,
Phys. Rev. C \textbf{101}, no.2, 024902 (2020)
[arXiv:1907.12786 [nucl-th]].

\bibitem{Abreu:2020jsl}
L.~M.~Abreu and F.~J.~Llanes-Estrada,
Eur. Phys. J. C \textbf{81}, no.5, 430 (2021)
[arXiv:2008.12031 [hep-ph]].

\bibitem{Albaladejo:2021cxj}
M.~Albaladejo, J.~M.~Nieves and L.~Tolos,
[arXiv:2102.08589 [hep-ph]].

\bibitem{Gamermann:2009fv}
D.~Gamermann and E.~Oset,
Phys. Rev. D \textbf{80}, 014003 (2009)
[arXiv:0905.0402 [hep-ph]].

\bibitem{Zhou:2017txt}
Z.~Y.~Zhou and Z.~Xiao,
Phys. Rev. D \textbf{97}, no.3, 034011 (2018)
[arXiv:1711.01930 [hep-ph]].

\bibitem{Li:2012cs}
N.~Li and S.~L.~Zhu,
Phys. Rev. D \textbf{86}, 074022 (2012)
[arXiv:1207.3954 [hep-ph]].

\bibitem{Junnarkar:2018twb}
P.~Junnarkar, N.~Mathur and M.~Padmanath,
Phys. Rev. D \textbf{99}, no.3, 034507 (2019)
[arXiv:1810.12285 [hep-lat]].

\bibitem{Faustov:2021hjs}
R.~N.~Faustov, V.~O.~Galkin and E.~M.~Savchenko,
Universe \textbf{7}, no.4, 94 (2021)
[arXiv:2103.01763 [hep-ph]].

\bibitem{ALICE:2015vxz}
J.~Adam \textit{et al.} [ALICE],
JHEP \textbf{03}, 081 (2016)
[arXiv:1509.06888 [nucl-ex]].

\bibitem{Loizides:2017ack}
C.~Loizides, J.~Kamin and D.~d'Enterria,
Phys. Rev. C \textbf{97}, no.5, 054910 (2018)
[erratum: Phys. Rev. C \textbf{99}, no.1, 019901 (2019)]
[arXiv:1710.07098 [nucl-ex]].

\bibitem{LHCb:2011zzp}
R.~Aaij \textit{et al.} [LHCb],
Eur. Phys. J. C \textbf{72}, 1972 (2012)
[arXiv:1112.5310 [hep-ex]].

\bibitem{CMS:2013fpt}
S.~Chatrchyan \textit{et al.} [CMS],
JHEP \textbf{04}, 154 (2013)
[arXiv:1302.3968 [hep-ex]].

\bibitem{LHCb:2020fvo}
R.~Aaij \textit{et al.} [LHCb],
JHEP \textbf{08}, 123 (2020)
[arXiv:2005.13422 [hep-ex]].

\bibitem{LHCb:2020sey}
R.~Aaij \textit{et al.} [LHCb],
Phys. Rev. Lett. \textbf{126}, no.9, 092001 (2021)
[arXiv:2009.06619 [hep-ex]].

\bibitem{Heijn:2728971}
Bo Heijn,
``The Measurement of the X(3872) production cross section via decays to $J/\psi\pi^+\pi^-$ in $pp$ collisions at $\sqrt{s} = 13$ TeV,''
Presented 14 May 2018,
http://cds.cern.ch/record/2728971

\bibitem{CMS:2011aqh}
S.~Chatrchyan \textit{et al.} [CMS],
JHEP \textbf{08}, 141 (2011)
[arXiv:1107.4800 [nucl-ex]].

\bibitem{ALICE:2013jfw}
E.~Abbas \textit{et al.} [ALICE],
Phys. Lett. B \textbf{726}, 610-622 (2013)
[arXiv:1304.0347 [nucl-ex]].
\end{thebibliography}
\end{document}